# The temperature-dependent non-Abelian gauge potential

Zheng-Chuan Wang

The University of Chinese Academy of Sciences, P. O. Box 4588,

Beijing 100049, China.

**Abstract**

This study shows that the usual time evolution operator can act as a U(1) gauge transformation on the initial wavefuntion. When considering the spin freedom in the system, the time evolution operator corresponds to a non-Abelian SU(2) gauge transformation for spin-1/2 particles and a SU(3) gauge transformation for spin-1 particles. If we adopt the adiabatic approximation in the magnetization dynamics of a ferromagnetic system driven by a spin-polarized current, a temperature-dependent non-Abelian thermal gauge potential will appear. We can employ a temperature-dependent Landau-Lifshitz-Gilbert (LLG) equation to describe the magnetization dynamics in the semi-classical limit, a linear approximated solution for this equation is given analytically.

E-Mail: wangzc@ucas.ac.cn

## I. Introduction

Yang-Mills gauge theory[1] is the foundation of elementary particle physics, which unifies the weak, electromagnetic, and strong interaction by the standard model. The fibre bundle theory in modern differential geometry manifests that the gauge potential corresponds to the connection in the principal-bundle[2], e.g., the gauge potential of gravity corresponds to the Cartan connection of Cartan verbein-bundle in space-time base manifold[2]. With the development of geometric phase theory[3], the gauge potential in the parameter space base manifold instead of the usual space-time manifold appears, and the integral of this gauge potential along a path in parameter space contributes to a geometric phase, which is in essence the element of the holonomy group associated with the fibre bundle[4]. The gauge potential is Abelian when the eigen-energy is non-degenerate, whereas it is non-Abelian when the eigen-energy is degenerate[5]. The gauge potential in the parameter space manifold plays an important role in topological physics, e.g., the topological energy band theory in the topological insulator[6]. There are other gauge potentials, such as the artificial or synthesized gauge potential, which appears in the man-made devices of electromagnetism and quantum optics[7]-a very active discipline; the emergent gauge potential which appears in the five-dimensional Kaluza-Klein theory and unifies Einstein's gravity theory and Maxwell's electromagnetic theory[8].

In 2025, we proposed a temperature-dependent thermal gauge potential based on the quantum Boltzmann equation (QBE), where a temperature-dependent damping force was derived from the scattering terms of QBE. We further relate this damping force with the U(1) thermal scalar and vector gauge potential, which provides a deep understanding of the microscopic expression of Luttinger's thermal scalar potential and Tatara's thermal vector potential[9]. This thermal damping force can be extended to the four-dimensional case, where the fourth component corresponds to a temperature-dependent power[10]. If we introduce the spin freedom into the QBE, we can obtain the so-called spinor Boltzmann equation (SBE)[11], which is a powerful tool for studying spin-polarized transport in spintronics. Similar to the case of QBE, the SBE can be exploited to derive a temperature-dependent spinor damping force from the scattering self-energy terms, which is a $2 \times 2$ matrix in spin space and can be related to a non-Abelian SU(2) thermal gauge potential as well as U(1) thermal gauge potential[12]. The usual SBE is only suitable to describe the transport of spin-1/2 particles, Sengstoch extended it to the case of spin-1[13], which can describe the spin-dependent transport of cold atom gases with big spin. A temperature-dependent spinor damp force was derived by us just as in the case of spin-1/2 , followed by a relation to the non-Abelian SU(3) thermal gauge potential and a U(1) thermal gauge potential[14]. Till now, both the Abelian and non-Abelian thermal gauge potentials had been explored by us in terms of the QBE and SBE.

Although we have established the connection between the thermal gauge potential and the temperature-dependent damping force as above, it remains an open question to determine completely the thermal gauge potential from the damping force. In addition to the framework of QBE and SBE, in 2026 a thermal gauge potential is obtained analytically based on the Born-Oppenheimer approximation[15], where the particles in the system vary slowly and the particles in the environment move quickly, a temperature-dependent thermal geometric phase naturally occurs by an integral of the thermal gauge potential. However, this thermal gauge potential is Abelian, not non-Abelian thermal gauge potential. If we further consider the spin freedom in the system, it is expected that a non-Abelian thermal gauge potential can be constructed analytically. In this manuscript, we explore this issue.

## II. Time evolution operator as a gauge transformation

Starting from a system with the Hamiltonian $H_0 = \frac{P^2}{2m} + V(x)$, where $\frac{P^2}{2m}$ is the kinetic energy, and $V(x)$ is the potential energy. The wavefunction at time $t$ evolves from the initial wavefunction at time $0$ by the time evolution operator $\psi(t) = exp(\frac{i}{\hbar} H_0 t)\psi(0)$, where $exp(\frac{i}{\hbar} H_0 t)$ is in fact an element of the U(1) group. As the time evolution operator depends on the position $x$ and time $t$, it can be regarded as a U(1) gauge transformation.

For a system with spin-1/2 particles, its Hamiltonian is $H_{1/2} = \frac{P^2}{2m} + V(x) + \frac{\hbar\mu_B}{2}\vec{B}(x)\cdot\vec{\sigma}$, where $\vec{B}(x)$ is the external magnetic field, $\vec{\sigma} = (\sigma_x, \sigma_y, \sigma_z)$ is the Pauli matrix, and $\mu_B$ the Bohr magnon. Its wavefunction experiences an evolution from the initial state as $\begin{pmatrix}\psi_\uparrow(t)\\ \psi_\downarrow(t)\end{pmatrix} = exp(\frac{i}{\hbar}H_{1/2}t)\begin{pmatrix}\psi_\uparrow(0)\\ \psi_\downarrow(0)\end{pmatrix}$. Since $H_{1/2}$ is a $2\times 2$ matrix, it can be expanded by the unit matrix $\hat{I}$ and the Pauli matrix $\vec{\sigma} = (\sigma_x, \sigma_y, \sigma_z)$ as $H_{1/2} = (\frac{P^2}{2m} + V(x))\hat{I} + \frac{\hbar\mu_B}{2}\vec{B}(x)\cdot\vec{\sigma}$. As $\sigma_x$, $\sigma_y$ and $\sigma_z$ are the generators of the SU(2) Lie group, in this way this time evolution operator can be regarded as a U(1)× SU(2) gauge transformation.

For a system with spin-1 particles, we can write its Hamiltonian as $H_1 = \frac{P^2}{2m} + V(x) + \mu_B\vec{B}(x)\cdot\vec{\hat{S}}$, where $\vec{\hat{S}} = (\hat{S}_x, \hat{S}_y, \hat{S}_z)$ is the spin operator, which is a $3\times 3$ matrix, we can decompose it by the unit matrix and 8 Gellman matrices $\hat{T}_{\mathrm{i}}(\mathrm{i} = 1...8)$ as $H_1 = (\frac{P^2}{2m} + V(x))\hat{I} + \frac{\hbar\mu_B}{2}\vec{B}(x)\cdot\vec{\hat{T}}$, where $\vec{\hat{T}} = (\hat{T}_1 ... \hat{T}_8)$ is the generator of the SU(3) Lie group, and the wavefunction feels the evolution as $\begin{pmatrix}\psi_1(t)\\ \psi_0(t)\\ \psi_{-1}(t)\end{pmatrix} = exp(\frac{i}{\hbar}H_1 t)\begin{pmatrix}\psi_1(0)\\ \psi_0(0)\\ \psi_{-1}(0)\end{pmatrix}$. In this respect the time evolution operator is a U(1)× SU(3) gauge transformation.

Except for the above case, we can study the system with other spins,

and the time evolution operator can still be viewed as a gauge transformation on the initial wavefunction. Overall, for a system without spin, it is a U(1) Abelian gauge transformation, and for a system with spin, it can be the non-Abelian gauge transformation.

## III. The temperature-dependent non-Abelian gauge potential

Let us investigate the magnetization dynamics of a ferromagnetic system driven by a spin transfer torque induced by a spin-polarized current. The Hamiltonian for the electrons in the ferromagnet is $h = \frac{P^2}{2m} + Ex + \frac{\hbar J}{2}\vec{M}(x)\cdot\vec{\sigma}$, where $\vec{\mathrm{E}}$ is the external electric field, $\vec{M}(x)$ is the magnetization of the ferromagnet, and $J$ is the exchange coupling constant for the s-d interaction of the spin-polarized electron and the local magnetic moment of the ferromagnet. The electronic wavefunction satisfies the following Schrödinger equation:

$$i\hbar\frac{\partial}{\partial t}\binom{\psi_\uparrow(t)}{\psi_\downarrow(t)} = \left(\frac{P^2}{2m} + Ex + \frac{\hbar J}{2}\vec{M}(x)\cdot\vec{\sigma}\right)\binom{\psi_\uparrow(t)}{\psi_\downarrow(t)}. \quad (1)$$

If we choose the direction of magnetization $\vec{M}$ as the z-axis, and write $\psi_{\uparrow,\downarrow}$ as $\psi_{\uparrow,\downarrow}(x) = R_{\uparrow,\downarrow}(x)exp(\frac{i}{\hbar}S_{\uparrow,\downarrow}(x))$, where $R_{\uparrow,\downarrow}(x)$ and $S_{\uparrow,\downarrow}(x)$ are all real functions. Substitute this expression into Eq.(1), and we have

$$i\hbar\frac{\partial}{\partial t}R_{\uparrow,\downarrow} - \hbar R_{\uparrow,\downarrow}\frac{\partial}{\partial t}S_{\uparrow,\downarrow} = \left(\frac{P^2}{2m} + Ex + \frac{\hbar J}{2}M_z\right)R_{\uparrow,\downarrow}. \quad (2)$$

Since the magnetization of the ferromagnet moves much slowly than the conduction electrons, we can adopt the adiabatic approximation to fix the magnetization $\vec{M}$ firstly and solve Eq. (1) for the conduction electrons, then substitute the s-d interaction into the equation of magnetization dynamics for the ferromagnet. As the conduction electrons move quickly than the magnetization, their spin will align parallel to the magnetization after a relaxation time because of the s-d exchange interaction between them, and finally arrive at an equilibrium state. Therefore we can describe the conduction electrons using the Dirac distribution function. The electronic density corresponding to energy $E_{\uparrow,\downarrow}$ is subject to $f_{\uparrow,\downarrow} = \frac{1}{exp[\frac{E_{\uparrow,\downarrow}}{k_B T(x,t)}]+1}$, where $T(x,t)$ is the temperature distribution in the system. On the other hand, the electronic density for energy $E_{\uparrow,\downarrow}$ can also be expressed by the wavefunction as $|\psi_{\uparrow,\downarrow}|^2$, if we let $f_{\uparrow,\downarrow} = |\psi_{\uparrow,\downarrow}|^2$, we can choose the amplitude of the wavefunction as $R_{\uparrow,\downarrow}(x) = \sqrt{\frac{1}{exp[\frac{E_{\uparrow,\downarrow}}{k_B T(x,t)}]+1}}$, which is similar to the method in [14]. Substitute it into Eq.(2). If the temperature doesn't change with time, we have

$$-\hbar \frac{\partial}{\partial t} S_{\uparrow,\downarrow} = \frac{(\frac{P^2}{2m}+Ex+\mu_B M)R_{\uparrow,\downarrow} - i\hbar\frac{\partial}{\partial t}R_{\uparrow,\downarrow}}{R_{\uparrow,\downarrow}}. \tag{3}$$

Finally, we obtain the action function $S_{\uparrow,\downarrow}$ as $S_{\uparrow,\downarrow} = -\frac{1}{\hbar}\int_0^t \frac{(\frac{P^2}{2m}+Ex+\mu_B M)R_{\uparrow,\downarrow} - i\hbar\frac{\partial}{\partial t}R_{\uparrow,\downarrow}}{R_{\uparrow,\downarrow}}$. This allows us to calculate the average $<\vec{s}> = <\psi_\uparrow, \psi_\downarrow|\vec{s}|\begin{matrix}\psi_\uparrow \\ \psi_\downarrow\end{matrix}>$ of the spin operator as $<\vec{s}> = 2R_\uparrow R_\downarrow cos(S_\uparrow - S_\downarrow)\vec{i} + 2iR_\uparrow R_\downarrow sin(S_\uparrow - S_\downarrow)\vec{j} + (R_\uparrow^2 - R_\downarrow^2)\vec{k}$, which is temperature-dependent because we introduce a temperature into the wavefunction under the adiabatic approximation. In addition to the spin-polarized electrons, the Hamiltonian for the local magnetic moment of the ferromagnet can be modeled as follows:

$$H_M = \mu_B \vec{B}_{eff}(x) \cdot \vec{\hat{S}} + J\mu_B < \vec{\hat{s}} > \cdot \vec{\hat{S}}, \quad (4)$$

where $\vec{\hat{S}}$ is the local spin operator of local moment. The time evolution operator $exp(\frac{i}{\hbar} H_M t)$ is a non-Abelian gauge transformation on the initial wavefunction. Since the s-d interaction term $J\mu_B < \vec{\hat{s}} > \cdot \vec{\hat{S}}$ between the conduction electron and the local moment is temperature-dependent, the gauge potential $\mu_B \vec{B}_{eff}(x) + J\mu_B < \vec{\hat{s}} >$ is a temperature-dependent thermal gauge potential, which is the focus of our manuscript.

Besides the Schrödinger and Heisenberg equation, LLG is another way to unveil the nature of magnetization dynamics in the semi-classical limit by inserting a spin transfer torque $J < \vec{\hat{s}} > \times \vec{M}$, which can be written as follows:

$$\frac{\partial}{\partial t} \vec{M} = -\gamma \vec{M} \times \vec{H}_{eff} + \alpha \vec{M} \times \frac{\partial}{\partial t} \vec{M} - J < \vec{\hat{s}} > \times \vec{M}, \quad (5)$$

which is temperature-dependent because of the spin average $< \vec{\hat{s}} >$. In [16], Zhang et al. proposed a generalized stochastic LLG equation with a spin transfer torque and investigated its thermal properties using the corresponding stochastic Fokker-Planck equation. However, in our temperature-dependent LLG equation, we straightforwardly introduce the temperature into the LLG equation using conduction electrons, which can be regarded as a reservoir with an equilibrium temperature under adiabatic approximation.

In the next section, we try to solve this temperature-dependent LLG equation approximately. At the stationary state $\frac{\partial}{\partial t}\vec{M} = 0$, Eq. (5) becomes to

$$\gamma\vec{M} \times \vec{H}_{eff} + J < \vec{\hat{s}} >\times \vec{M} = 0, \qquad (6)$$

so the stationary solution is $\vec{M}_0(\mathrm{x}) = \gamma\vec{H}_{eff} + J < \vec{\hat{s}} >$. The general solution has a deviation away from the stationary solution, and can be written as $\vec{M}(x,t) = \vec{M}_0(\mathrm{x}) + \delta\vec{M}(x,t)$. If the magnetic oscillation has a frequency $\omega$, the time dependence of $\delta\vec{M}(x,t)$ can be simplified as $\delta\vec{M}(x,t) = \vec{A}(x)e^{i\omega t}$,, where $\vec{A}(x)$ is the amplitude function to be determined. Substituting this into Eq. (5), we have

$$i\omega\vec{A}(x)e^{i\omega t} = (\gamma\vec{H}_{eff} + J\mu_B < \vec{\hat{s}} >) \times \vec{A}(x)e^{i\omega t} + i\omega\alpha(\vec{M}_0(\mathrm{x}) + \vec{A}(x)e^{i\omega t}) \times \vec{A}(x)e^{i\omega t}. \qquad (7)$$

If the deviation $\delta\vec{M}(x,t)$ is small, we can neglect the small higher nonlinear term $i\omega\alpha\vec{A}(x)e^{i\omega t} \times \vec{A}(x)e^{i\omega t}$ and obtain a linear equation for $\vec{A}$ as follows:

$$i\omega\vec{A}(x) = (\gamma\vec{H}_{eff} + J\mu_B < \vec{\hat{s}} >) \times \vec{A}(x) + i\omega\alpha\vec{M}_0(\mathrm{x}) \times \vec{A}(x). \qquad (8)$$

Its characteristic value is determined by setting the determinant to zero

$$\begin{vmatrix} iw & B_z & -B_y \\ -B_z & iw & B_x \\ B_y & -B_x & iw \end{vmatrix} = 0, \text{ where } \vec{B} = \gamma\vec{H}_{eff} + J < \vec{\hat{s}} > + i\omega\alpha\vec{M}_0(\mathrm{x}),$$

resulting in the three characteristic values $\omega_1 = 0$, $\omega_\pm = \pm\sqrt{(B_x^2 + B_y^2 + B_z^2)}$ with corresponding characteristic vectors $\vec{A}_1 =$

$$(\frac{B_x}{B_z}, \frac{B_y}{B_x}, 1),\ \vec{A}_{\pm} = (\frac{B_y(i\omega_{\pm} + \frac{B_x B_z}{B_y})}{B_z(1 - \frac{i\omega_{\pm}}{B_z})(B_z + \frac{i\omega_{\pm} B_x}{B_y})}, \frac{B_y}{B_z} - \frac{i\omega_{\pm} B_y(i\omega_{\pm} + \frac{B_x B_z}{B_y})}{B_z^2(1 - \frac{i\omega_{\pm}}{B_z})(B_z + \frac{i\omega_{\pm} B_x}{B_y})}, 1),$$

respectively. Since $< \vec{\hat{s}} >$ in Eq.(8) is temperature-dependent, the linear solution of magnetization is temperature-dependent, too.

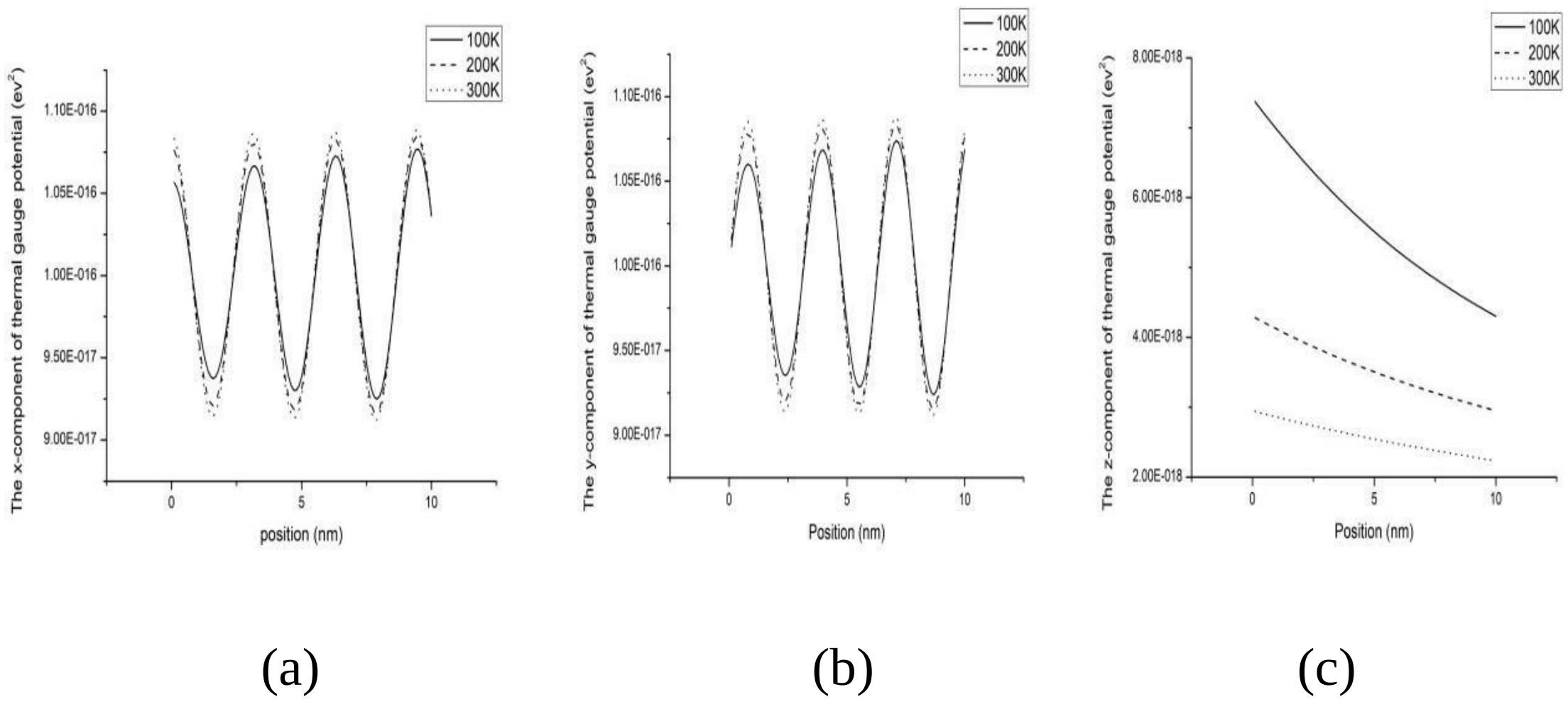

(a) (b) (c)

Fig.1 The three components of the thermal gauge potential at different temperatures of 100K, 200K, and 300K: (a) the x-component (b) the y-component (c) the z-component, where the temperature gradient is adopted as 1K/nm.

In Fig.1, we numerically demonstrate the temperature-dependent gauge potential $\mu_B \vec{B}_{eff}(x) + J < \vec{\hat{s}} >$ at different temperatures of 100K, 200K, and 300K, where the external magnetic field is set to a constant. We can see that this thermal gauge potential changes with temperature slightly, which is due to the system we choose, in other systems the change in thermal gauge potential may not be so small. In the ferromagnetic system driven by a spin-polarized current, when the temperature varies from 100 to 300K, the variation of energy caused by the thermal fluctuation is $\sim\ k_B T = 8.617 \times 10^{-5} eVK^{-1} \times 200K = 0.0172eV$, while the s-d interaction for the spin-polarized electron and the local magnetic moment is about $0.1eV$. Thus the energy of thermal fluctuation is smaller than the exchange energy. The variation of thermal gauge potential by temperature is allowed to be small in Fig.1. Maybe we can choose other perfect systems which suffer from the obvious change in thermal gauge potential.

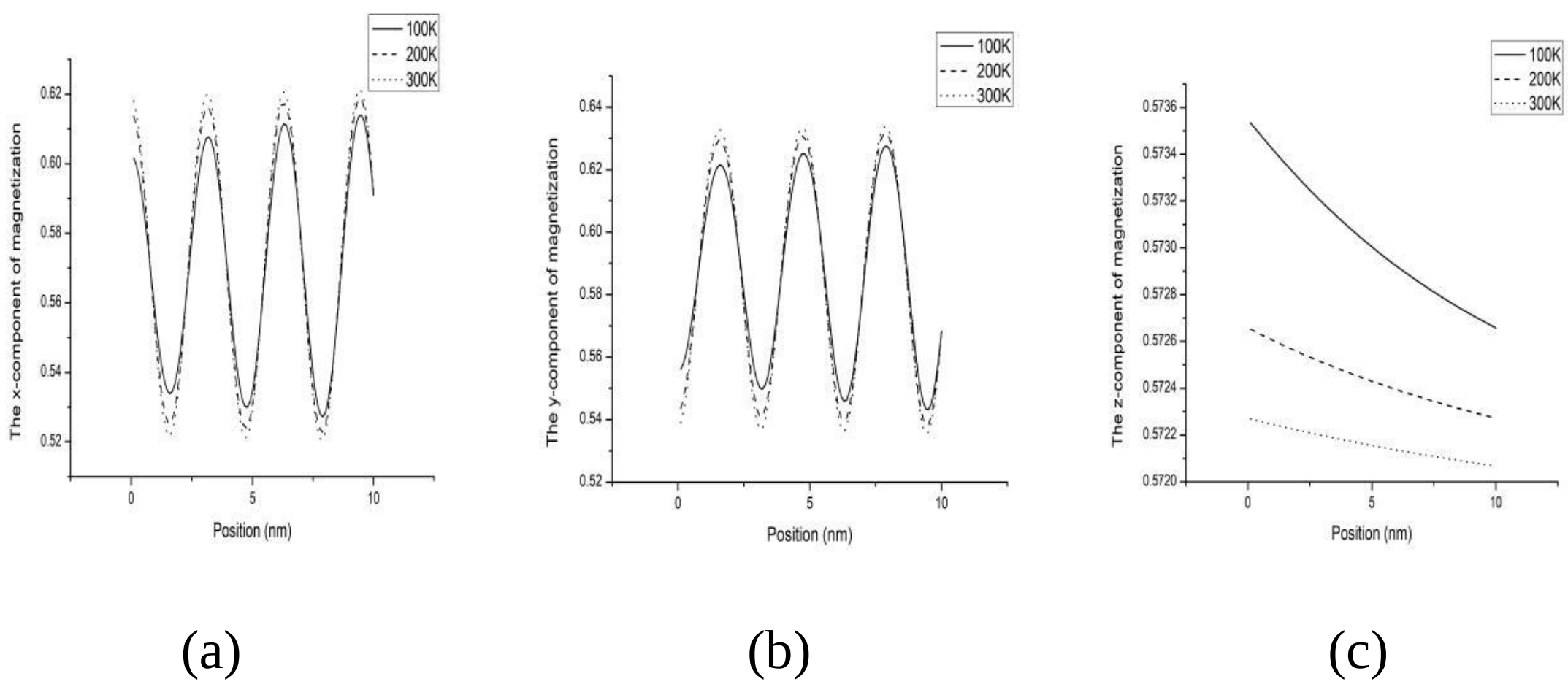

(a) (b) (c)

Fig.2 The three components of the magnetization at different temperatures of 100K, 200K, and 300K: (a) the x-component (b) the y-component (c) the z-component, where the temperature gradient is adopted as 1K/nm.

The solution of magnetization as a function of position at different temperatures of 100K, 200K, and 300K for characteristic value $\omega_1 = 0$ is shown in Fig.2. It exhibits that the temperature influences magnetization, which offers important clues to investigate the temperature-dependence of the LLG equation. The x- and y-components oscillate as a function of position with increasing amplitudes, which is due to the temperature increases from the left to the right boundary in the system with a temperature gradient. When the temperature increases, the temperature-dependent amplitude of the wavefunction increases, leading to an increase in the oscillation amplitude. The z-component decreases with position, which is a sharp contrast to the x-and y-components, implying the difference in position-dependence among different components. The shape of the curves in Fig.2 is similar to the corresponding curves in Fig.1, the s-d exchange interaction between the spin-polarized electrons and the local magnetization is responsible for this. The solution of magnetization for the other two characteristic values $\omega_{\pm} = \pm\sqrt{(B_x^2 + B_y^2 + B_z^2)}$ is similar to Fig.2, but we don't show them here.

## IV. Summary and discussions

In summary, we have demonstrated that the time evolution operator can be viewed as a gauge transformation on the initial wavefunction. For a system without spin, the gauge transformation is Abelian, whereas it is non-Abelian for the system with spin. When we explore the magnetization dynamics of a ferromagnet driven by a spin-polarized current, we can obtain a temperature-dependent thermal gauge potential under an adiabatic approximation, and the corresponding non-Abelian gauge transformation is temperature-dependent. The temperature is introduced by the conduction electrons, which arrive at an equilibrium state with a temperature after a relaxation time. In the semi-classical limit, we obtain a temperature-dependent LLG equation to explore the temperature-dependence of magnetization dynamics, its linear approximate solution is shown numerically in Fig. 2, which changes with temperature obviously.

Note that the above gauge potential and thermal gauge potential are different from the Berry connection concerning the topological magnetic texture in ferromagnet[17], e.g., the magnetic Skyrmion which has recently attracted intensive interest. For the Skyrmion, there are the Berry connection and curvature along with its magnetic texture, the integral of the Berry curvature will contribute a topological charge to characterize this Skyrmion, whereas its Berry connection is a U(1) Abelian gauge potential, which is induced by the slow variation of the magnetic texture. It is different from the Abelian and non-Abelian gauge potentials in our manuscript, which originate from the interaction in the Hamiltonian. For example the thermal gauge potential comes from the s-d exchange interaction of conduction electrons and the local magnetic moment of the ferromagnet. Certainly, it is interesting to explore the temperature-dependent Berry connection for the magnetic texture of the Skyrmion, it can be obtained straightforwardly by substituting the expression of temperature-dependent magnetization into the Berry connection formula for magnetic texture.

**Acknowledgments**

This study is supported by the National Key R&D Program of China (Grant No. 2022YFA1402703).

**Data Availability Statement**

Data sets generated during the current study are available from the corresponding author on reasonable request.

**Additional information**

Competing interest statement: The authors declare that they have no competing interests.